\documentclass[aps,twocolumn,preprintnumbers,prd,superscriptaddress,10pt]{revtex4-1}
\makeatletter
\def\set@curr@file#1{%
  \begingroup
    \escapechar\m@ne
    \xdef\@curr@file{\expandafter\string\csname #1\endcsname}%
  \endgroup
}
\def\quote@name#1{"\quote@@name#1\@gobble""}
\def\quote@@name#1"{#1\quote@@name}
\def\unquote@name#1{\quote@@name#1\@gobble"}
\makeatother
\usepackage{amssymb,amsmath,amsthm,amsfonts}
\usepackage[usenames]{color}
 
\usepackage{xcolor}
\usepackage{bm}
\usepackage{dcolumn}
\usepackage[utf8]{inputenc}
\usepackage{latexsym}
\usepackage{rotating}
\usepackage{longtable}

\usepackage{enumerate}
\usepackage{tensor,multirow}
\usepackage{url}
\usepackage[linktocpage]{hyperref}
\usepackage{float}
\usepackage{graphicx}

\def\be{\begin{equation}}
\def\ee{\end{equation}}
\newcommand{\beq}{\begin{eqnarray}}
\newcommand{\eeq}{\end{eqnarray}}
\begin{document}

\title{The hole picture}


\author{Vitor~Cardoso}
\affiliation{CENTRA, Departamento de F\'{\i}sica, Instituto Superior T\'ecnico
  -- IST, Universidade de Lisboa -- UL, Avenida Rovisco Pais 1, 1049 Lisboa,
  Portugal\\
  Email: vitor.cardoso@ist.utl.pt}
\begin{abstract}
Because of the very definition of black holes --- no light escapes them and falling objects get infinitely faint when approaching --- it is impossible to ever prove that they exist. However, electromagnetic and gravitational-wave observatories have now `seen' black holes. Datasets from these observations, released in 2019 and late 2018, give important hints about the environment, origin and growth of black holes. 
%
%
%
\end{abstract}

\maketitle

100 years ago, Newton's theory of gravitation was (almost literally) eclipsed by general relativity, when
the British expeditions to Princ\'{i}pe island and Sobral measured how light of distant stars was deflected as it
passed close to the Sun. For this kind of phenomenon, which occurs in the solar system and on human timescales, general relativity predicts but a small correction to Newton's theory. However, general relativity bore three unexpected offspring: black holes (BHs), gravitational waves and spacetime singularities. These have been at the heart of the most interesting developments in theoretical and experimental physics of the last decades. 
The most precise observations ever have `seen' the first two of these offspring last year, overcoming atmospheric turbulence, the faintness of such objects and various noise sources. These breakthroughs may be the beginning of the journey to understanding how to quantize gravity and perhaps ``cure'' the pathological behavior at singularities; perhaps they are the first steps towards the ultimate description of gravity.

The concept of BHs dates to 1915. While on the war front, Karl Schwarzschild found a solution to the equations of general relativity that describes the vacuum left behind a collapsing, non-spinning star. Spin was included much later, in 1963 by Roy Kerr. These objects are now known as BHs and are very special indeed: they are the only solution of general relativity describing stationary, isolated collapsed objects. In the words of Subrahmanyan Chandrasekhar in his 1983 Nobel lecture,
``the only elements in the construction of black holes are our basic concepts of space and time. They are, thus, almost by definition, the most perfect macroscopic objects there are in the universe.'' 
The clean and simple appearance of a BH is due to its horizon, where spacetime is so warped that it truly cuts a hole in the universe. Although the collapsed progenitor star that forms the BH gives rise to pathological regions in the BH interior, where physical quantities blow up or physics stops being deterministic~\cite{Reall:2018}, from the outside, as seen by distant observers, time stops at the horizon and light sources approaching it fade completely. Anything that passes the horizon cannot come out. Observers are thus unable to probe events beyond this veil.

\begin{figure}[ht!]
\includegraphics[width = 245pt]{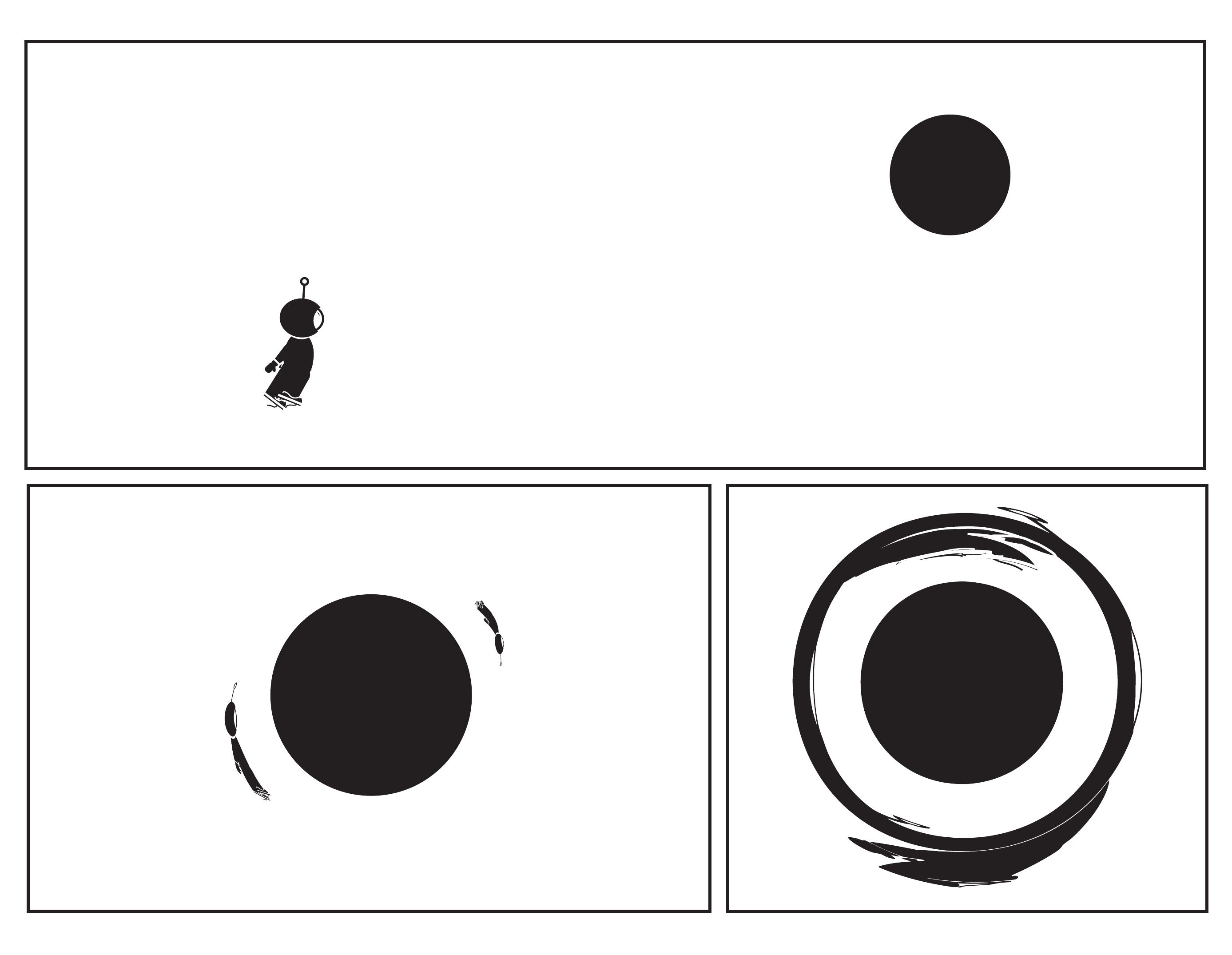}
\caption{
\textbf{`Seeing' a black hole.} \textbf{a$\vert$} An astronaut roams in space and drifts by a black hole (BH). \textbf{b$\vert$} Initially, an observer sees both a direct and a secondary image of the astronaut, corresponding to light rays that `went around' the BH. \textbf{c$\vert$} When the astronaut is directly behind the BH, he is seen from all angles, and his image forms an Einstein-like ring, like that observed by the Event Horizon Telescope. The central black circle corresponds to all photons that fall into the BH. Used \href{https://www.esa.int/gsp/ACT/phy/Projects/Blackholes/WebGL.html}{{ESA ACT BH visualization}}.
%
\label{BH_image}}
\end{figure}

For those not brave enough to jump into a BH, the sole means of learning more about these objects is to study their exterior. At large distances, objects follow Newtonian dynamics; planets can orbit BHs on stable, circular trajectories, for example. However, very close to the BH, spacetime is warped, altering drastically the dynamics of light and matter. Stable motion of matter is no longer possible, and the material simply falls into the BH. Whereas the deflection of light by the sun, as measured by the 1919 expeditions, is tiny, close to BHs light deflection can be enormous (Fig.~\ref{BH_image}). In fact, light can even turn around. The geometric nature of spacetime acts like a lens, creating distorted and sometimes secondary, or even higher order images~\cite{Luminet:2019hfx}. 

To test the above predictions, one needs a BH and an astronaut close to it, or some luminous matter instead. 
Supermassive BHs are the perfect target: theory predicts that they sit at the center of galaxies, where they sink due to `dynamical' friction with the rest of the galaxy. Theory also predicts that remains of disrupted stars and interstellar material should orbit these BHs. Friction between this material heats it up, and it becomes bright and therefore visible.

Observations conform to the theory. In late 2018 and 2019, two spectacular measurements were done, with unprecedented accuracy, using different telescopes \cite{Akiyama:2019cqa,GRAVITY}. 
Using new techniques (optical/infrared interferometry and radio very large baseline interferometry)
astronomers were able to obtain outstanding resolutions. These are equivalent to seeing a Euro coin on the Moon, from Earth. 

The first experiment, GRAVITY, combined the four 8-meter telescopes of the European Southern Observatory to form a super-telescope with equivalent diameter of 130 m, with unprecedented combination of angular resolution and sensitivity.
The team monitored in this way, at wavelengths of the order of a micrometer, the centre of our galaxy, where a 4 million solar-mass BH lurks, and studied orbiting lumps of luminous material extremely close to the BH, in fact as close as possible without falling in~\cite{GRAVITY}. These flares were moving at roughly $30\%$ of the speed of light, and their motion was consistent with all predictions of general relativity.

Simultaneously, a 6 billion solar mass dark object at the centre of the M87 galaxy was being scrutinized by another set of telescopes: the first ever image of a BH was announced by the Event Horizon Telescope (EHT) project~\cite{Akiyama:2019cqa}. The team gathered light from orbiting and ejected material around M87, and performed a very careful comparison with a multitude of astrophysical scenarios, varying the inclination of the orbit, magnetic fields, BH mass and spin among other factors, to determine the conditions that fit observations (and several do). The final `image' was released in April 2019 and is indeed spectacular: a dark central core is surrounded by a luminous rim, analogous to that in Fig.~\ref{BH_image}.

These results make it possible to finally start understanding the astrophysical conditions close to BHs, the physics of accretion and magnetic fields, what type of material is orbiting the BH and where it came from. 
Although only a beginning, other fundamental questions can also be posed in the context of these observations. 
The observations from GRAVITY and EHT can be used to start placing strong constraints on the nature of the object~\cite{Cardoso:2019rvt}. The observation of the polarization of light from the dark object by the EHT can exclude the presence of light-polarizing particles~\cite{Chen:2019fsq}; the mass and spin estimates exclude some ``superradiant'' scenarios which arise in dark matter models. For example, if photons were massive, they could spin BHs down very rapidly. Thus the observation of massive BHs can be used for particle physics~\cite{Davoudiasl:2019nlo,Bar:2019pnz}!

These electromagnetic observations are now complemented by a radically new way of doing astrophysics. In 2015, humankind opened a new channel with which to understand the universe: gravitational waves, ripples in spacetime that travel at the speed of light. These waves are produced by accelerated motions, and are emitted copiously when BHs collide. 
These waves are detected by monitoring miniscule changes in the relative lengths of two L-shaped kilometre-length arms of a laser interferometer. 

Gravitational waves are magnificent probes of the cosmos because, unlike light, they interact feebly with matter. They travel freely through the universe, and like eternal entities they roam, retaining the shape they had when generated. Gravitational waves thus offer a unique glimpse of dynamical BHs and compact stars. In late 2018, the LIGO--Virgo collaboration released the catalogue of all the events seen during the first and second observation runs of the detectors (up to September 2017)~\cite{LIGOScientific:2018mvr}. There are 11 events, one of them being the collision between two neutron stars, the other 10 involving BHs. The BHs have a mass between 10 and 50 solar masses, and the collisions can release the equivalent of 3 solar masses in energy, in a fraction of a second. These are amazingly powerful and violent collisions. The LIGO--Virgo catalogue transmits two important messages: the detections use BH templates and general relativity, and all that everything seen to date is consistent with what Einstein's equations predict. 
The BHs that were seen with gravitational waves are more massive that those detected in X-ray emitting binaries, 
which begs for an explanation --- how were they born, and how did they grow? But most importantly, the detectors are being upgraded. Thus, future generations of gravitational-wave detectors will see better, farther and more events. The release of new gravitational-wave detections by the LIGO--Virgo collaboration is expected anytime soon, so stay tuned (quite literally: there are real-time applications that notify when detections are made).

BHs exist throughout the universe. The latest observations, in the electromagnetic spectrum and in the gravitational-wave band, have given the strongest evidence for their existence to date. Astronomers can `see them' and study their 
physics to exquisite precision. These observations promise to teach when, how and where BHs are born and how they grow, and what type of environment they live in.
Given the puzzles associated with BHs~\cite{Reall:2018,Compere:2019ssx}, it is only natural to dream that unexpected features will manifest in future observations, perhaps bringing a step closer a more consistent description of gravity and of all interactions.

\vskip 2mm
\noindent \textcolor{black}{\bf Key Advances:}

\noindent \textcolor{darkgray}{$\bullet$ GRAVITY combines four telescopes to form the equivalent of a telescope of 130 m diameter. It has observed luminous material extremely close to the black hole at the centre of our galaxy.}

\noindent \textcolor{darkgray}{$\bullet$ The Event Horizon Telescope combines radio telescopes in different continents, equivalent to a detector as big as our planet. This has been used to observe the surroundings of the black hole at the centre of the Messier 87 galaxy, revealing a dark core surrounded by a luminous halo.}

\noindent \textcolor{darkgray}{$\bullet$ The latest full Ligo--Virgo data release contains 11 events, 10 of which involve black holes.}


\noindent \textcolor{darkgray}{$\bullet$ All observations are consistent with general relativity and predictions for black holes, but the coming years will bring more insights.}

\vskip 2mm

\noindent{\bf{\em Acknowledgments.}}
I am thankful to Ana Sousa for the image and for discussions. I am indebted to E. Berti, P. Cunha, K. Destounis, Frank Eisenhauer, M. C. Ferreira, J. Grover, C. Herdeiro, D. Hilditch, J. P. S. Lemos and Frederic Vincent, for important feedback.


\end{document}